\documentclass[12pt]{article}

\usepackage{graphicx,amssymb,url,mathrsfs, amsmath}
\usepackage{eucal}
\usepackage{wrapfig}
\usepackage{boxedminipage}
\usepackage{setspace}
\usepackage{subfigure}
\usepackage[dvipdfm]{hyperref}

\def\IR{{\hbox{{\rm I}\kern-.2em\hbox{\rm R}}}}
\def\IB{{\hbox{{\rm I}\kern-.2em\hbox{\rm B}}}}
\def\IN{{\hbox{{\rm I}\kern-.2em\hbox{\rm N}}}}
\def\IC{\,\,{\hbox{{\rm I}\kern-.59em\hbox{\bf C}}}}
\def\IZ{{\hbox{{\rm Z}\kern-.4em\hbox{\rm Z}}}}
\def\IP{{\hbox{{\rm I}\kern-.2em\hbox{\rm P}}}}
\def\IH{{\hbox{{\rm I}\kern-.4em\hbox{\rm H}}}}
\def\ID{{\hbox{{\rm I}\kern-.2em\hbox{\rm D}}}}

\def\be{\begin{equation}}
\def\ee{\end{equation}}
\def\ba{\begin{eqnarray}}
\def\ea{\end{eqnarray}}

\def\ea{{\it et al}. }

\begin{document}

\begin{titlepage}

\begin{flushright}

\end{flushright}
\vspace{0.5in}

\begin{center}
{\large \bf Improved AdS/QCD Model with Matter}\\
\vspace{10mm}
Alexander Stoffers and Ismail Zahed\\
\vspace{5mm}
{\it Department of Physics and Astronomy, Stony Brook University, Stony Brook NY 11794}\\
          \vspace{10mm}
{\tt \today}
\end{center}
\begin{abstract}
We study an improved AdS/QCD model at finite temperature and chemical potential. An Ansatz for the $\beta$-function for
the boundary theory allows for the derivation of a charged dilatonic black hole in bulk.  The solution is asymptotically
RN-AdS in the UV and AdS$_2 \times \mathbb{R}^3$ in the IR. We discuss the thermodynamical aspects of the solution.
The fermionic susceptibilities are shown to deviate from the free fermionic limits at asymptotic temperatures despite the
asymptotically free nature of the gauge coupling at the boundary. The Polyakov line, the temporal and spatial string
tensions dependence on both temperature and chemical potential are also discussed.
\end{abstract}
\end{titlepage}

\renewcommand{\thefootnote}{\arabic{footnote}}
\setcounter{footnote}{0}



\section{Introduction}
Non-perturbative QCD at finite chemical potential is still elusive. First principle formulations
such as the lattice suffer from the sign problem~\cite{SIGN}. Most non-perturbative formulations based
on semiclassics such as the instanton or dyon formulations require further insights on the
role of the fermionic zero modes at finite chemical potential. Some insights on the role of
the chemical potential at finite temperature can be gleaned from strong coupling
lattice QCD~\cite{DEFORCRAND} or models~\cite{MODELS}.
Most of these models for light 2-flavor QCD suggest a second order transition at
small chemical potential and finite temperature, and a first order transition at higher
chemical potential. Noteworthy is the occurence of a tricritical point which appears to be
sensitive to the nature of the confining forces.

Non-perturbative QCD with a large number of colors at finite chemical potential is likely
a crystal of confined baryons.  The crystal binding energy can be parametrically small,
causing it to melt under quantum fluctuations and/or temperature. The result is a strongly
coupled baryonic liquid. The original Skyrme model supports this descriptive although it
suffers from the inherent shortcomings of the higher order chiral terms at high density~\cite{SKYRME}.
The chiral holographic approaches to QCD support the crystal structure in the large
number of  colors limit without the shortcomings of the Skyrme model~\cite{HOLO}.
The crystal is found to melt at relatively small temperatures by the
Lindemann criterion, resulting into a holographic liquid of instantons at low density
and dyons at higher densities. Therefore, it is reasonable to think that the cold and
dense inhomogeneous phase gives way to a homogeneous phase under the effects
of temperature.

The purpose of this work is to address issues in relation to the dense
and hot homogeneous baryonic phase using the (bottom-up) holographic approach.
Specifically, we will use a variant of the improved (bottom-up) holographic approach
to finite temperature Yang-Mills in the large number of colors limit put forward by Gursoy et al.
\cite{Gursoy:2007cb, Gursoy:2007er, Alanen:2009xs} with the addition of a fermion chemical
potential to account for baryon number. In~\cite{Gursoy:2007cb, Gursoy:2007er}
a potential for the dilaton field is constructed to reproduce some key features of Yang-Mills theory in
4 dimensions, namely heavy quark confinement in the infrared and asymptotic freedom in the
ultraviolet. Recently, a variant of this construction was suggested in~\cite{Alanen:2009xs}
whereby the dilaton field is directly tied to the running of the gauge coupling in the Yang-Mills
theory. This construction is more transparent physically as it directly ties the holographic direction
to the running of the gauge coupling constant. It is also less numerically intensive.
We follow~\cite{Alanen:2009xs} and add a U(1) charge field in bulk to account for the
effects of a finite chemical potential at the boundary.

In section 2 we discuss the bottom-up improved holographic model in 5 dimensions
with bulk gravity coupled to a dilaton and the U(1) charged field. The dilaton dynamics
follows from the running of the gauge coupling. The dilaton potential is fixed by the
equations of motion. Explicit solutions are constructed in section 3. In section 4 we discuss the
small and big charged black hole solutions, and their ensuing bulk thermodynamics.
In section 5 we analyze the fermionic susceptibilities and compare them to the ideal gas
limit. In section 6, the Polyakov line is evaluated at finite temperature and chemical potential.
In section 7 we discuss the role of the chemical potential on the spatial and temporal
string tension. Our conclusions are in section 8.  The sensitivity of the model to the 
dilaton-gauge-field coupling is discussed in Appendix A. In Appendix B we derive 
(quark) susceptibilities for a warped holographic model without a dilaton.

\section{The Model}
In the Einstein frame the 5-dimensional Einstein-Maxwell action in the background of a scalar or dilaton field $\phi$, is given by \cite{Gursoy:2007cb, Gursoy:2007er}
\be
S=\frac{1}{16 \pi G_5}\left(\int d^5x \sqrt{-g} \Big( R -\frac{4}{3} \partial_A\phi\partial^A\phi+V(\phi)-\frac{\mathcal{L}^2}{4} e^{c \phi}F_{AB}F^{AB} \Big)  -2\int d^4x \sqrt{h} K \right )
\label{action}
\ee
with the $U(1)$ gauge field tensor $F_{AB}=\partial_{[A}A_{B]}$ and $\mathcal{L}$ the radius of the AdS-space in the conformal limit. The Gibbons-Hawking boundary term gives no contribution to the equations of motion but is crucial for evaluating the on-shell action \cite{Gursoy:2008za}. The coupling to the $U(1)$ charge is a generalization of \cite{Gubser:2009qt}, where we consider the parameter $c$ in the exponential gauge coupling as a free parameter. A recent analysis in this direction is found in \cite{Charmousis:2010zz}. For a non-constant scalar field $\phi=\phi(z)$ the Bianchi identity $\nabla^{A}G_{AB}=0$ (with $G_{AB}$ the Einstein tensor) ensures that the Einstein equations imply the equation of motion for the scalar field. With the following Ansatz for the metric
\be
ds^2=b(z)^2\left(-f(z) dt^2+d\vec{x}^2+\frac{1}{f(z)}dz^2 \right)
\label{metric}
\ee
we follow \cite{Gursoy:2007cb, Gursoy:2007er, Alanen:2009xs} and assume that the $\beta$-function for the gauge field theory on the boundary at $z=0$ is given by
\be
\beta(\lambda)=b{d\lambda\over db}=-\beta_0\lambda^q
\label{1st}
\ee
with the running t'Hooft coupling $\lambda(z)  = e^{\phi(z)} \sim  g_{YM}^2 N_c$. Note that (\ref{1st}) does not follow from varying the action (\ref{action}). $q \geq 1$ ensures confinement in the IR and the values $q={10}/{3}$ and $\beta_0=488.8$ have been shown in \cite{Alanen:2009ej} to reproduce the dilaton potential in \cite{Gursoy:2009jd} to lowest order in $\lambda$. We will use these values in our numerical analysis.
For a static, charged black hole solution the equation of motion for the scalar potential $A_0(z)$ is given by
\begin{eqnarray} \nonumber
b(z)^{-5} \partial_z \left( b(z) e^{c \phi(z)} A_0'(z) \right) \propto \delta (z-z_{charge}) \\
A_0'(z)=-\frac{\textbf{e}}{\mathcal{L}^3}\frac{e^{-c \phi(z)}}{b(z)}
\label{eomA}
\end{eqnarray}
with $A_0'(z)=\frac{\partial}{\partial z}A_0(z)$ and the electric charge $\textbf{e}$ located at some position $z_{charge}$ behind the horizon of the black hole. We can now solve the Einstein equations
\begin{eqnarray}
6\frac{b'^2}{b^2}-3\frac{b''}{b}-\frac{4}{3}\phi'^2=0 \label{2nd} \\
\frac{3}{2}\frac{b'f'}{b}+\frac{1}{2}f''-\frac{\mathcal{L}^2 e^{c\phi}}{2 b^2}A_0'^2=0  \label{3rd} \\
\frac{9}{2}\frac{b'f'}{b f}+3\frac{b''}{b}+6\frac{b'^2}{b^2}+\frac{1}{2}\frac{f''}{f}-\frac{b^2}{f} V=0 \label{4th}
\end{eqnarray}
together with (\ref{1st}) and (\ref{eomA}). Note that the factor $f(z)$ drops out from the spatial components of the energy-momentum tensor for the gauge field.

\section{Solutions}
The solution for the electrostatic potential is given by
\be
A_0(z)=-\frac{\textbf{e}}{\mathcal{L}^3} \int_0^z dx \frac{e^{-c\phi(x)}}{b(x)} + \mu \ .
\ee
The integration constant $\mu$ is interpreted as the chemical potential since $A_0(z \rightarrow 0)=\mu$ \cite{Kim:2006gp}.
The requirement for the electrostatic potential to vanish at the horizon of the black hole, see (\ref{Temp}), gives
\be
A_0(z_H)=0 \Leftrightarrow \mu=\frac{\textbf{e}}{\mathcal{L}^3} \int_0^{z_H}dx\frac{e^{-c\phi(x)}}{b(x)} \ .
\label{electrostaticpotential}
\ee
The warping factor $b(z)$ is not influenced by the presence of the charge and the solution to (\ref{1st}) is given by
\be
Q=\ln{\frac{b}{b_0}}=\frac{1}{(q-1)\beta_0} \frac{1}{\lambda^{q-1}} \ .
\label{Q(lambda)}
\ee
Since
\be
W(\lambda)=\frac{-b'(z)}{b(z)^2}=W(0)e^{\frac{a}{4Q}}
\ee
and $a=\Big(\frac{4}{3(1-q)}\Big)^2$, (\ref{2nd}) yields
\be
z=\frac{\mathcal{L}}{b_0}\int_Q^{\infty} dx \ e^{-x-\frac{a}{4 x}} \ , \label{z(Q)}
\ee
where the constant $1/W(0)=\mathcal{L}$ is fixed by the boundary value of the potential $V(z)$, see
(\ref{V(Q)}) and \cite{Alanen:2009xs}. (\ref{Q(lambda)}) and (\ref{z(Q)}) imply that neither the warping $b(z)$, nor the coupling $\lambda$ depend on the temperature or chemical potential. ($\ref{z(Q)}$) introduces a scale $\Lambda = \frac{b_0}{\mathcal{L}}$ for our model and implies an upper bound on the radial coordinate
\be
z\leq \frac{\mathcal{L}}{b_0}\int_0^{\infty} dx e^{-x-\frac{a}{4 x}}=z_w \ . \label{zw}
\ee
A static wall in the IR is generated which leads to an area law for the Polyakov loop as we will detail below.

Expanding the integral in $(\ref{z(Q)})$ for small $z$ (large $Q$)
\begin{eqnarray}
\int_Q^{\infty} dx e^{-x-\frac{a}{4 x}} &=& \sum_0^\infty {1\over n!}\left(-\frac{a}{4}\right)^n \Gamma(1-n,Q) \\
&=&\exp\left[-Q-{a\over 4Q}\right]\left(1+{a\over 4Q^2}-{a\over 2Q^3} \ + \ ...\right)
\end{eqnarray}
with the incomplete Gamma-function $\Gamma(1-n,Q)$ we obtain $b(z \rightarrow 0) = \frac{\mathcal{L}}{z}$ and $\phi(z\rightarrow 0) \propto -\log{\left(-\log{\Lambda z}\right)}$, which reproduces a logarithmic running coupling for the model on the boundary.\\
Changing variables from $z$ to $Q$ in (\ref{3rd}) reads
\be
\partial_Q^2 f(Q) + \left(-\frac{a}{4Q^2}+4\right)  \partial_Q f(Q) = \frac{((q-1)\beta_0)^{\frac{c}{q-1}}\textbf{e}^2}{\mathcal{L}^2 b_0^6} e^{\frac{-a}{2Q} -6Q}Q^{\frac{c}{q-1}} \ .
\ee
The two asymptotic conditions on $f(z)$, $f(z=0)=f(Q=\infty)=1$ and $f(Q_H)=0$ at the horizon $Q_H=Q(z_H)$, fix the two integration constants and we obtain
\begin{eqnarray}
f(Q)= 1 + C_1 i(Q) + \delta  \frac{(\mu / \Lambda)^2}{j(Q_H)^2}\int_{\infty}^{Q} dx \ e^{-\frac{a}{4x}-4x} \int_{\infty}^x dy \ e^{-\frac{a}{4y}-2y} y^{\frac{c}{q-1}}
\end{eqnarray}
with $\delta=\frac{1}{((q-1)\beta_0)^{\frac{3c}{1-q}}}$ and
\begin{eqnarray}
C_1&=&\frac{
-1-\delta \frac{(\mu / \Lambda)^2}{j(Q_H)^2}\int_{\infty}^{Q_H} dx \ e^{-\frac{a}{4x}-4x} \int_{\infty}^x dy \ e^{-\frac{a}{4y}-2y} y^{\frac{c}{q-1}}}{i(Q_H)} \\
j(Q)&=& \int_{\infty}^{Q}dx \ e^{-\frac{a}{4x}-2x} x^{\frac{c}{1-q}}\\
i(Q)&=&\int_{\infty}^{Q}dx \ e^{-\frac{a}{4x}-4x}.
\end{eqnarray}
We see that our solutions for $f(z)$ reduces to the one obtained in \cite{Alanen:2009xs} for $\mu=0$. With a vanishing coupling of the dilaton to the gauge field, i.e. $c=0$, the metric in the Einstein frame (\ref{metric}) approaches the RN-AdS-metric for $z \rightarrow 0$ with
\be
f(z \rightarrow 0) \rightarrow 1 - \frac{C_1}{4} \left( \Lambda z \right)^4 + \frac{(\mu / \Lambda)^2}{24 j(Q_H)^2} \left( \Lambda z \right)^6 \ .
\ee
We can now solve (\ref{4th}) for the scalar potential
\begin{eqnarray}
V(Q)=W(0)^2 e^{\frac{a}{2Q}}\Big(\frac{1}{2} \partial_Q^2 + \big(5-\frac{a}{8 Q^2} \big)\partial_Q + 12-\frac{3a}{4 Q^2} \Big) f(Q) \label{V(Q)}
\end{eqnarray}
As $\phi$ is temperature independent this implies $V(Q)=V(\phi, T, \mu)$. The dependence on temperature and chemical potential
follows from $f(Q)$.
The normalization $V(z\rightarrow 0)=V(Q\rightarrow \infty)=\frac{12}{\mathcal{L}^2}$ gives $\frac{1}{W(0)}=\mathcal{L}$.

\section{Thermodynamics}
The Hawking temperature of the black hole,
\be
T=\frac{1}{4 \pi}f'(z_H) \ , \label{Temp}
\ee
is given by
\be
T=\frac{-\Lambda}{4 \pi}\frac{e^{-3 Q_H}}{i(Q_H)} \left(1- \delta (\mu / \Lambda)^2 \frac{k(Q_H)}{j(Q_H)^2} \right) \label{temperature}
\ee
with $z_H, Q_H=Q(z_H)$ the position of the horizon and
\be
k(Q)=\int_{\infty}^{Q} dx e^{-\frac{a}{4x}-4x} \int_x^{Q} dy e^{-\frac{a}{4y}-2y} y^{\frac{c}{q-1}}.
\ee

\begin{figure}[!htbp]
  \begin{center}
  \includegraphics[width=8cm]{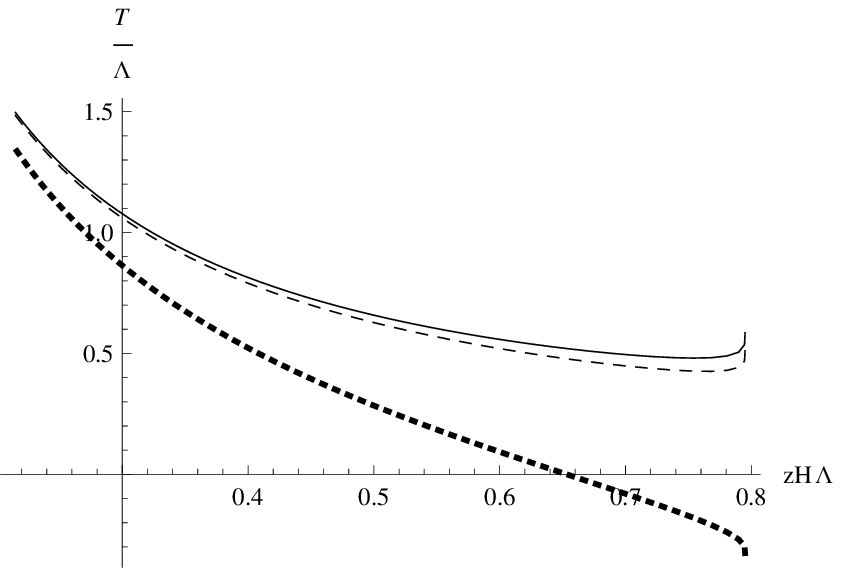}
  \includegraphics[width=8cm]{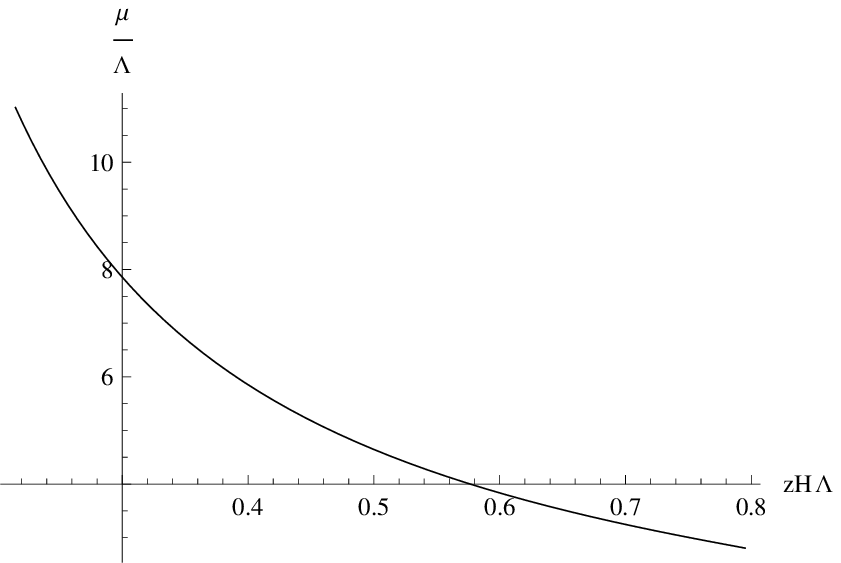}
  \caption{Left: The Hawking temperature at chemical potentials $\mu / \Lambda=0 \ \text{(solid)}, \ 1 \ \text{(dashed)}$, \ 3.5 \ \text{(dotted)} as a function of the scaled horizon $z_H \Lambda$. Right: The chemical potential at $T=0$. ($c=0$)}
  \label{figuretemperature}
  \end{center}
\end{figure}

At vanishing chemical potential the black hole solution has a minimal non-zero temperature. The two regions $T'(z_H)>0$ $(T'(Q_H)<0)$ and $T'(z_H)<0$ $(T'(Q_H)>0)$ correspond to a small and a big black hole branch. Since the solution for $\mu=0$ shows a minimum temperature, a certain density ($\mu_0= 2.8 \Lambda$) is needed to obtain a solution with zero temperature. Here, we will focus on the case of vanishing dilaton-gauge-field coupling, i.e. $c=0$, and consider the case $c=4$ separately in Appendix A. For $T \rightarrow 0$ and $\mu \geq \mu_0$, the small black hole vanishes leaving only the big black hole solution, see Fig. \ref{figuretemperature}, and the metric in the IR becomes AdS$_2 \times \mathbb{R}^3$, \cite{Lu:2009gj, Faulkner:2009wj}. With $(z-z_H)= \gamma / \zeta$, $t= 2/(\gamma f''(z_H)) \tau$ and $\gamma \rightarrow 0$, $\zeta, \tau$ finite, (\ref{metric}) reduces to
\be
ds^2=\left(\frac{2b(z_H)^2}{f''(z_H)}\right) \frac{1}{\zeta^2} \left(- d\tau^2+d\zeta^2\right) + b(z_H)^2 d\vec{x}^2 \ .
\ee
One crucial step in the above analysis is that the temperature as a function of the horizon does not have a minimum as $T \rightarrow 0$, i.e. $f''\left(z_H(T=0,\mu)\right) \neq 0$. In our case this translates to the disappearance of the small black hole as $T \rightarrow 0$.

Note that regularizing the action in (\ref{action}) by subtracting a 'vacuum' (thermal gas) solution with functions $\phi_0, b_0, f_0=1$
does not necessarily yield the grand potential since the scalar potential (\ref{V(Q)}) has additional temperature and chemical potential
dependence thereby upsetting the Gibbs relations. This  notwithstanding, we can still define
the entropy density $s(T,\mu)$ carried by the black hole through its area as

\be
s(T,\mu)=\frac{b_0^3}{4 G_5} e^{3 Q_H}.
\ee
The small (big) black hole branches have negative (positive) specific heat, $c_v \propto T ds/dT$, and are unstable (stable).

\begin{figure}[!htbp]
  \begin{center}
  \includegraphics[width=8cm]{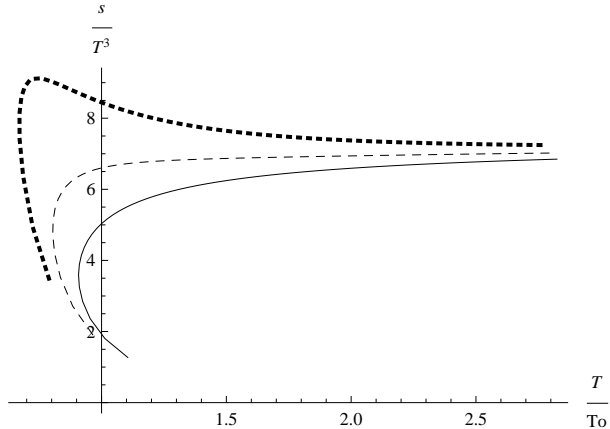}
  \caption{The scaled entropy density $s/T^3$ at chemical potentials $\mu / \Lambda=0 \ \text{(solid)}, \ 1 \ \text{(dashed)}$, \ 1.5 \ \text{(dotted)} with $T_0=170$MeV and $\Lambda=321$MeV. Both the contributions from the small and big black hole are shown.}
  \label{figureentropyc0}
  \end{center}
\end{figure}

\begin{figure}[!htbp]
  \begin{center}
  \includegraphics[width=8cm]{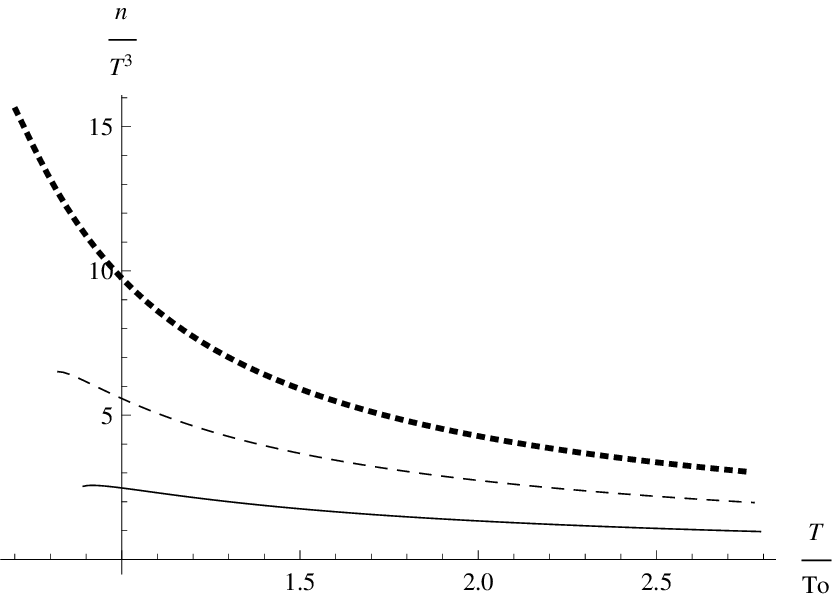}
  \includegraphics[width=8cm]{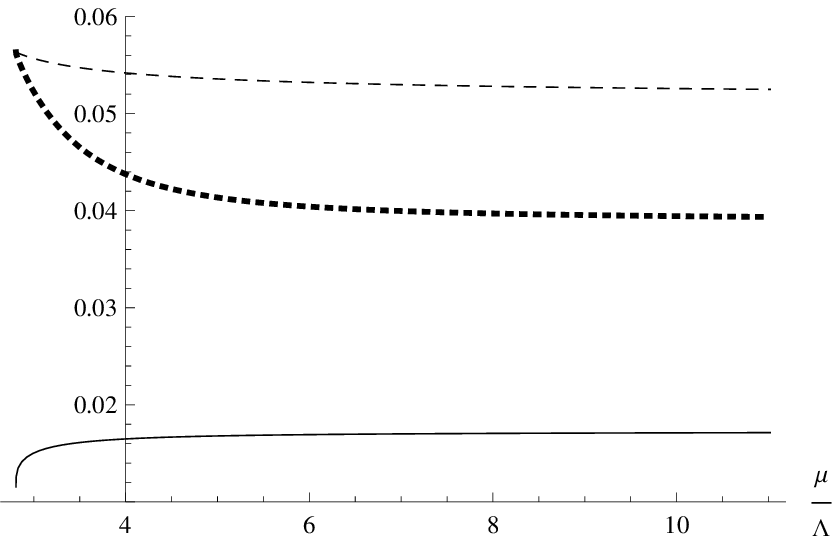}
  \caption{Left: Scaled charge density $n/T^3$ for the big black hole at densities $\mu / \Lambda =0.5 \ \text{(solid)}, \ 1 \ \text{(dashed)}, \ 1.5 \ \text{(dotted)}$ with $\alpha= \frac{3}{2 \pi^2} \Lambda^4$. Right: $s/\mu^3$ (solid), $n/\mu^3$ (dashed), $e/\mu^4$ (dotted) at $T=0$.}
  \label{figuredensity}
  \end{center}
\end{figure}

The scaled entropy density $s/T^3$ reaches its asymptotic value around $T\simeq 2 T_0$ and develops a peak with increasing $\mu$. A low temperature and high density expansion yields 
\begin{eqnarray}
s(\mu \gg T)=\frac{b_0^3}{4 G_5} \left( \frac{1}{6\sqrt{6} }\frac{\mu^3}{\Lambda^3} +\frac{\pi}{4} \frac{\mu^2 T}{\Lambda^3} + \frac{\pi^2 3 \sqrt{3}}{8 \sqrt{2}} \frac{\mu T^2}{\Lambda^3}+  \frac{a\pi \sqrt{3}}{16 \sqrt{2}} \frac{\mu T^2}{\Lambda^3 \log{\frac{\mu}{\Lambda \sqrt{6}}}} + \mathcal{O}(T/\mu)\right).
\label{ST}
\end{eqnarray}
(\ref{ST}) reduces to the RN-AdS entropy~\cite{Sin:2009rf} up to logarithmic corrections.
Much like the RN-AdS black hole, our black hole solution carries a finite entropy at zero temperature
and a linear specific heat, see Fig. \ref{figuredensity}. The effects of the running coupling through the scalar field causes
only logarithmic corrections much like the high temperature case.

The pressure at vanishing chemical potential is obtained by integrating the entropy density over the small and big black hole branch
\be
p(T,\mu=0)=\int_0^{Q_H(T,\mu=0)} dQ \left( \frac{\partial T}{\partial Q} \right) s(Q(T,\mu=0)).
\label{pressureszeromu}
\ee
The scale $\Lambda=\frac{b_0}{\mathcal{L}}$ is fixed by requiring that the pressure at $\mu=0$ vanishes at a critical temperature $T_0=170$MeV or $\Lambda=321$MeV. Comparison with an ideal gluon gas  gives  $\frac{b_0^3}{G_5}=\frac{16 N_c^2 \Lambda^3}{45 \pi}$ \cite{Alanen:2009xs}.

\begin{figure}[!htbp]
  \begin{center}
  \includegraphics[width=8cm]{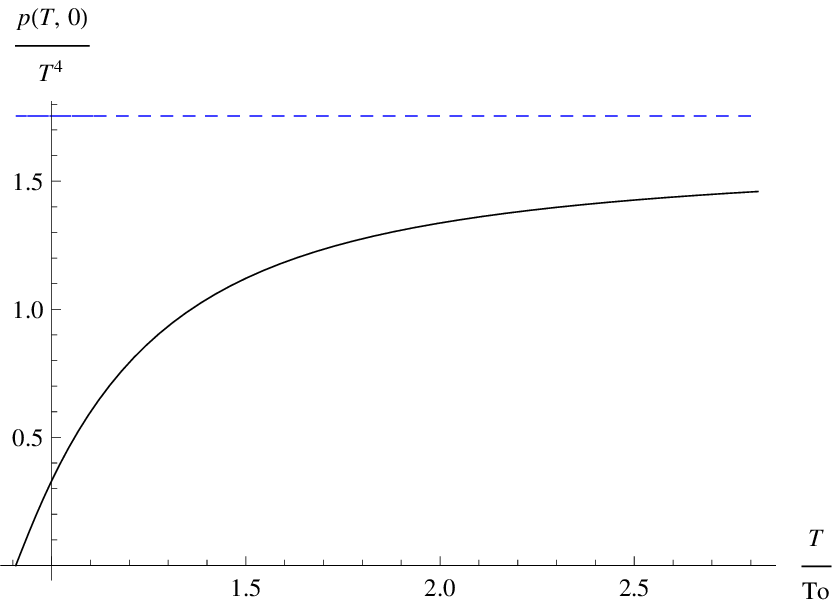}
  \includegraphics[width=8cm]{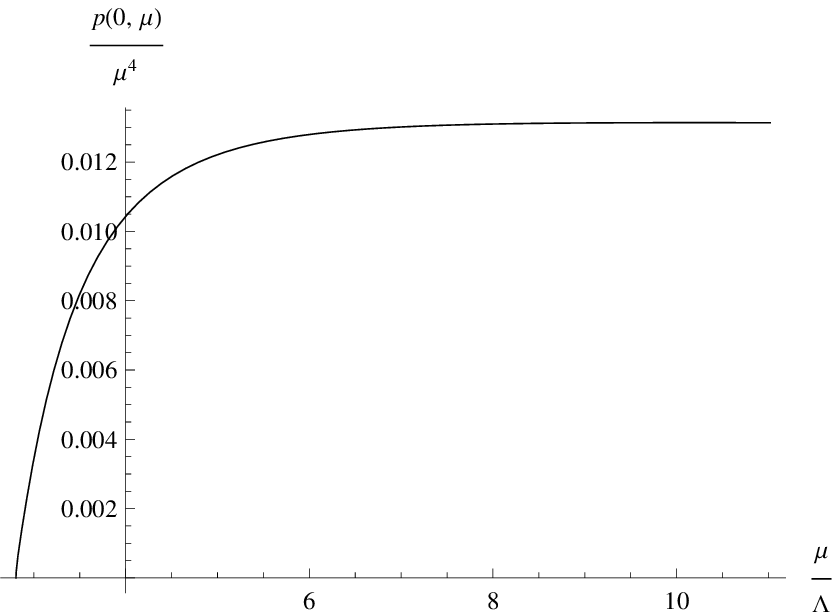}
  \caption{Left: Big black hole pressure at $\mu=0$ ($N_c=3$) and pressure of an ideal gluon gas $P_{ideal \ gas}/T^4=2\left(N_c^2-1\right)\frac{\pi^2}{90}$ (dashed). Right: Pressure at $T=0$ ($N_c=N_f=3$). For an ideal relativistic Fermi gas of $N_f$ massless quark flavors $P_{ideal \ gas}/\mu^4=\frac{N_f}{4 \pi^2}$.}
  \label{figurepressurezeromu}
  \end{center}
\end{figure}

The pressure of the small black hole is negative, indicating its instability. Fig. \ref{figurepressurezeromu} shows the big black hole contribution to the pressure. For $T/T_0 < 1$ the 'vacuum' solution dominates.
The integration constant in (\ref{pressureszeromu}) is fixed by analyzing the high temperature thermodynamics of the small black hole, \cite{Gursoy:2008za}. Since the small black hole vanishes for $T=0$, we fix the integration constant by setting the zero temperature pressure to zero at $\mu_0=\mu_{min}\Big|_{T=0}=\mu\left(T=0,Q_H=0\right)=2.8 \Lambda$,
\be
p(T=0,\mu)=\int_0^{Q_H(T=0,\mu)} dQ \left( \frac{\partial \mu}{\partial Q} \right) n(Q(T=0,\mu)),
\label{pressureszeroT}
\ee
with the charge density $n$ given by (\ref{density}). Fig. \ref{figurepressurezeromu} shows that the black hole solution is unstable against the vacuum solution below $\mu_0$. In contrast, the RN-AdS black hole at zero temperature \cite{Sin:2007ze}, dominates the action for all chemical potentials down to $\mu=0$.

For fixed chemical potential the charge density $n(T,\mu)$ is proportional to the electric charge $\textbf{e}$ and using (\ref{electrostaticpotential}) we write
\be
n = \frac{\alpha}{b_0^4}\textbf{e}=\frac{\alpha \mathcal{L}^3 \mu}{b_0^4 \int_0^{z_H}dx\frac{e^{-c\phi(x)}}{b(x)}}=\frac{-\alpha \delta^{-1/3} \mu}{\Lambda^2 m(Q_H)}
\label{density}
\ee
with the proportionality constant $\alpha$ and
\be
m(Q)=\int_{\infty}^{Q} dx Exp\left(-\frac{a}{4x}-2x\right) x^{\frac{c}{q-1}} \ .
\ee
For $\mu \gg T$ we obtain
\begin{eqnarray}
n(\mu \gg T) = \frac{2 \alpha}{\Lambda} \left( \frac{1}{6} \frac{\mu^3}{\Lambda^3} (1+\frac{a}{4 \log{\frac{\mu}{\Lambda \sqrt{6}}}}) + \frac{\mu^2 T}{\Lambda^3} (\frac{\pi }{\sqrt{6}} + \frac{a \pi (3+\sqrt{6})}{24\log{\frac{\mu}{\Lambda \sqrt{6}}}}) + \mathcal{O}(T/\mu) \right).
\end{eqnarray}
As in the RN-AdS, the leading order low temperature correction is linear in $T$. The deviation from the pure RN-AdS shows up as
logarithmic corrections.

\section{Susceptibilities}
In QCD at low density and high temperature the pressure can be expanded as
\be
\frac{p(T,\mu)}{T^4}= \sum^{\infty}_{n=0}c_n(T)\left(\frac{\mu}{T}\right)^n
\label{pressureexpansion}
\ee
with the (flavor symmetric) quark susceptibilities
\be
c_n(T) = \frac{1}{n!} \frac{\partial^n}{\partial(\frac{\mu}{T})^n} \frac{p(T,\mu)}{T^4} \Big|_{\mu=0}.
\ee
Various hadronic susceptibilities in the transition region around $T_0 = 170$MeV at vanishing quark chemical potential show distinct characteristics.
In particular, lattice studies \cite{Allton:2005gk} and expectations from PNJL models \cite{Ghosh:2006qh} both confirm that $c_4, c_6$ show distinct
peaks at a critical temperature $T_c$. Asymptotics of the susceptibilities come close to the ideal gas values at temperatures $T \simeq 2 T_c$. These charge density fluctuations are obtained as derivatives with respect to the density on the grand canonical partition function.

The  quark number susceptibility in hard and soft wall AdS/QCD models were studied in \cite{Kim:2006ut, Kim:2010zg}. In these models the first non-vanishing coefficient in the expansion (\ref{pressureexpansion}) show a jump at a critical temperature due to a Hawking-Page transition.
In the improved model under consideration with running gauge coupling, we can explicitly assess the first few moments  in (\ref{pressureexpansion}).
We first note that the odd moments vanish since $m(Q_H)$ in (\ref{density}) receives corrections at high $T$ (high $Q_H$) of the form $\mu^2/T^2$. From (\ref{density}) we obtain
\begin{eqnarray}
c_2(T)&=&\frac{1}{2T^2}\frac{\partial}{\partial \mu} n(T,\mu) \Big|_{\mu=0}\\
&=& \frac{- \alpha \delta^{-1/3}}{2T^2 \Lambda^2}  \frac{1}{m(Q_H)}\Big|_{\mu=0} \ .
\label{c2}
\end{eqnarray}
Note that eqn. (\ref{Temp}) implies
\be
\left(\partial Q_H \over \partial \mu \right)\Big|_{\mu=0}=0 \ ,
\ee
\begin{eqnarray}
\left(\partial^2 Q_H \over \partial \mu^2 \right)\Big|_{\mu=0}&=&\frac{\delta }{2 \pi \Lambda  T }\frac{Exp\left(-3Q_H\right)}{\left[3 i(Q_H)+ Exp(-4Q_H-\frac{a}{4Q_H})\right]}\frac{k(Q_H)}{(j(Q_H))^2} \Big|_{\mu=0} \ ,\\
\left(\partial^4 Q_H \over \partial \mu^4 \right)\Big|_{\mu=0}&=&-3 \left(\partial^2 Q_H \over \partial \mu^2 \right)^2 \left(\frac{\partial^2 T}{\partial Q_H^2}\right) \left(\frac{\partial Q_H}{\partial T}\right) \Big|_{\mu=0} .
\end{eqnarray}
We obtain for $c_4(T), c_6(T)$
\begin{eqnarray}
c_4(T)&=&\frac{1}{24} \frac{\partial^3}{\partial \mu^3} n(T,\mu) \Big|_{\mu=0} \\
&=& \frac{-\alpha \delta^{-1/3}}{24 \Lambda^2}\left(\partial^2 Q_H \over \partial \mu^2 \right)\frac{\partial}{\partial Q_H} \frac{1}{m(Q_H)}\Big|_{\mu=0} \\
c_6(T)&=&\frac{T^2}{6!}\frac{\partial^5}{\partial \mu^5} n(T,\mu) \Big|_{\mu=0}\\
&=&
\frac{-\alpha \delta^{-1/3} T^2}{6! \Lambda^2} \left( \Big(\frac{\partial^4 Q_H}{\partial \mu^4}\Big) \frac{\partial}{\partial Q_H} + 3\Big(\frac{\partial^2 Q_H}{\partial \mu^2}\Big)^2 \frac{\partial^2}{\partial Q_H^2}  \right) \frac{1}{m(Q_H)}\Big|_{\mu=0}.
\label{c4c6}
\end{eqnarray}
As can be seen from a high temperature expansion of the integrals in (\ref{c2}), (\ref{c4c6}) ($Q_H \rightarrow \infty$) $c_4$ and $c_6$ do not converge to a finite asymptotic high temperature value \emph{unless} $c=0$. Thus, a meaningful comparison to QCD demands a vanishing coupling of the dilaton $\phi$ to the $U(1)$ charge. With $c=0$ the high temperature asymptotics are given by
\begin{eqnarray}
c_2 &\rightarrow&  \pi^2 \left(\frac{\alpha}{\Lambda^4}\right) \\ \label{susceptibilitiesasymptotics}
c_4 &\rightarrow& \frac{1}{18}\left(\frac{\alpha}{\Lambda^4}\right) \\
c_6 &\rightarrow& \frac{1}{540 \pi^2} \left(\frac{\alpha}{\Lambda^4}\right).
\end{eqnarray}

\begin{figure}[!htbp]
  \begin{center}
  \includegraphics[width=8cm]{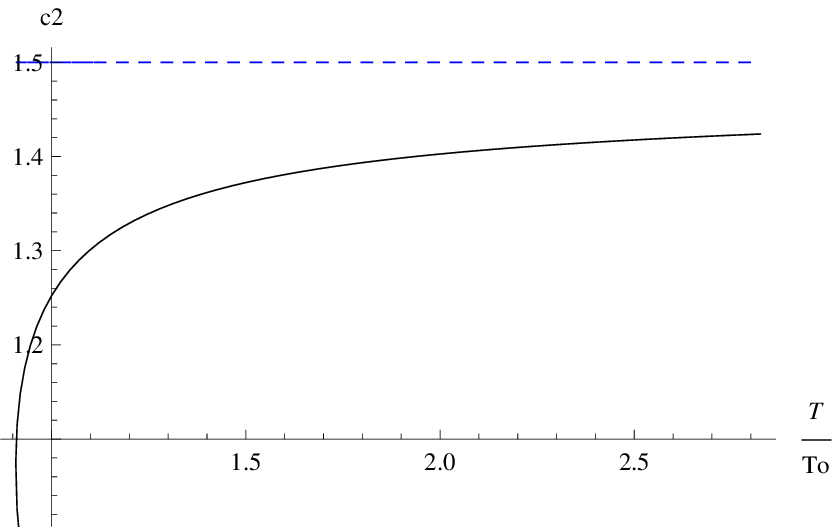}
  \includegraphics[width=8cm]{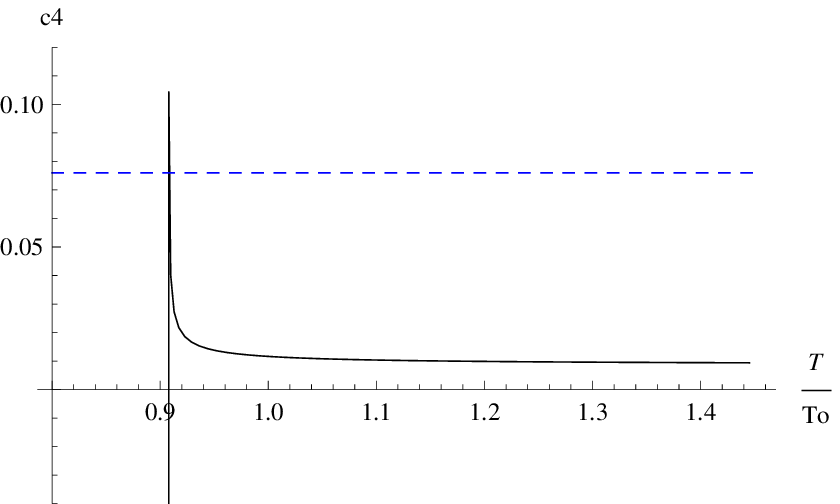}
  \caption{Susceptibilities for the big black hole. Dashed line: ideas gas value for $N_c=N_f=3$.}
  \end{center}
\end{figure}

\begin{figure}[!htbp]
  \begin{center}
  \includegraphics[width=8cm]{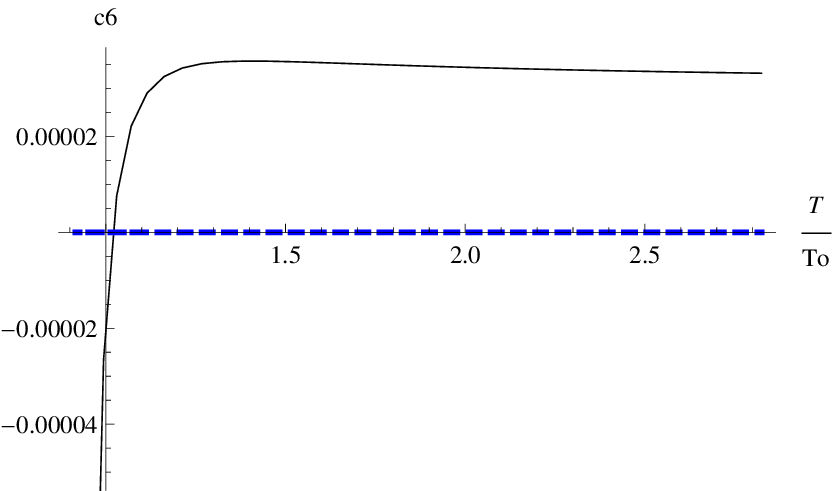}
  \includegraphics[width=8cm]{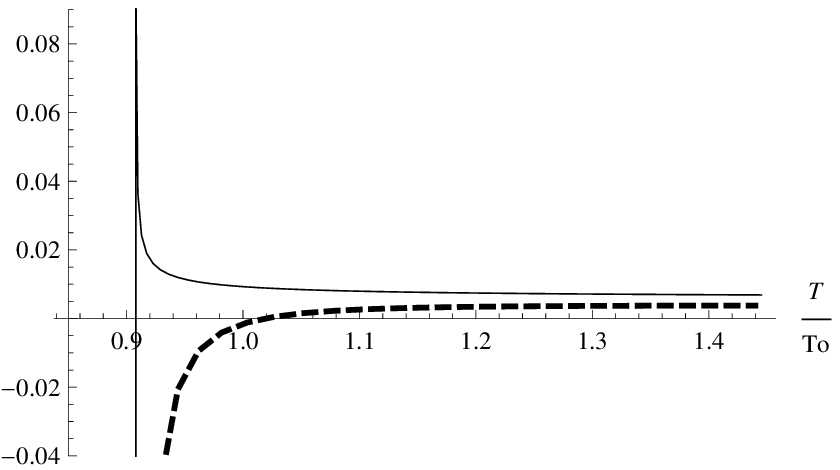}
  \caption{Left: $c_6$ and ideal gas value for $N_c=N_f=3$ (dashed). Right: Ratios $c_4/c_2$ (solid) and $c_6/c_4$ (dashed)}
  \end{center}
\end{figure}

As a result of the non-trivial warping $b(z)$, the susceptibilities are non-constant and a comparison with recent lattice data \cite{Allton:2005gk} shows that the susceptibilities $c_2, c_4, c_6$ obtained in this model have the correct shape around the critical temperature $T_0$. $c_2$ and $c_6$ approach its asymptotic value from below, while $c_4$ shows a distinct peak at the critical temperature. All high temperature asymptotics of the susceptibilities are strictly positive. For the model proposed in \cite{Andreev:2006nw}, the susceptibilities vanish as explained in Appendix B.\\

In comparison, for an ideal gas of massless quarks and anti-quarks the susceptibilities read $c_2^{ideal \ gas}=\frac{N_c N_f}{6} $, $c_4^{ideal \ gas}=\frac{N_c N_f}{12 \pi^2}$
, $c_6^{ideal \ gas}=0$ with the ratios $c_4^{ideal \  gas}/c_2^{ideal \ gas}= 1/(2 \pi^2)$, $c_6/c_4=0$. In our model: $c_4/c_2=1/(18 \pi^2)$, $c_6/c_4=1/(30 \pi^2)$.
While the scale $\Lambda$ was fixed by the gluonic part of the pressure, we fix the constant $\alpha$ by comparing to the fermionic part of the pressure
of an ideal gas. For $N_c=3$, $N_f=3$ we fit the high temperature asymptotic of $c_2$ to the ideal gas value and obtain $\alpha= \frac{3}{2 \pi^2} \Lambda^4$.
The susceptibilities in the holographic model considered in \cite{Sin:2009rf} with a non-confining RN-AdS black hole are constant and given by
$c_2 ^{RN-AdS}=N_c^2 \gamma^2 /8$, $c_4^{RN-AdS}= \frac{N_c^2 \gamma^4}{48 \pi^2}$, $c_6^{RN-AdS}=-\frac{N_c^2 \gamma^6}{216 \pi^4}$, where the flavor dependence is embedded in the parameter $\gamma^2 \propto N_f/N_c$.\\

\begin{figure}[!htb]
  \begin{center}
  \includegraphics[width=8cm]{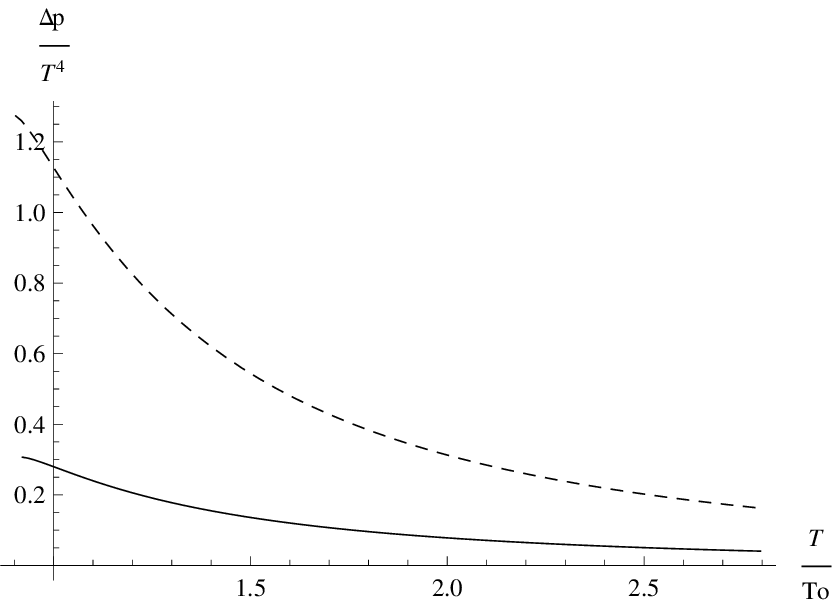}
  \includegraphics[width=8cm]{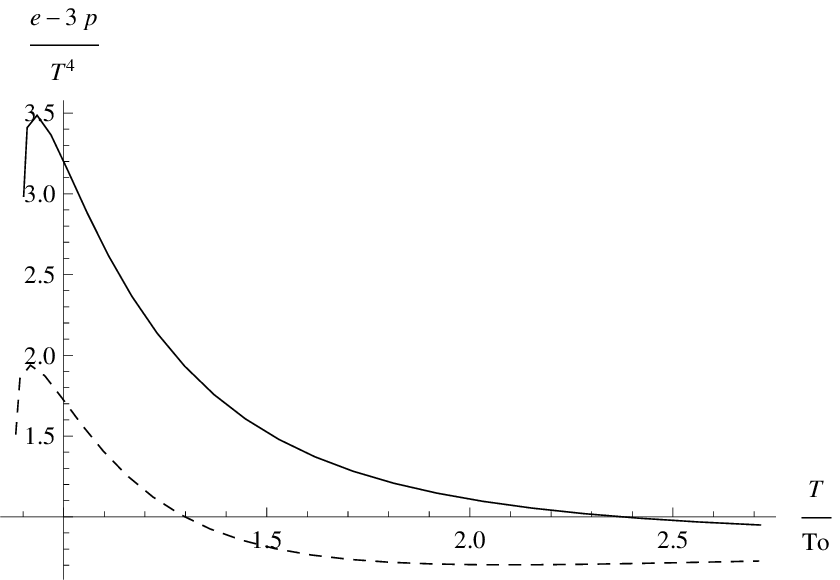}
  \caption{$\Delta p/T^4$ and $(e-3p)/ T^4$ for $\mu / \Lambda = 0.25  \ \text{(solid)}, \ 0.5 \ \text{(dashed)}$.}
  \label{figuredeltap}
  \end{center}
\end{figure}

It is now straightforward to evaluate the density contribution to the pressure, $\Delta p(T,\mu\neq 0)/T^4=\sum^{\infty}_{n=2}c_n(T)\left(\frac{\mu}{T}\right)^n \simeq c_2(T)\left(\frac{\mu}{T}\right)^2+c_4(T)\left(\frac{\mu}{T}\right)^4+c_6(T)\left(\frac{\mu}{T}\right)^6 $, and the energy density $e(T, \mu)=T s-p+\mu n$ at high temperatures and small charge density. Fig. \ref{figuredeltap} shows our results for $(e-3p)/T^4$ at different densities.

\section{Polyakov line}
\begin{figure}[!htbp]
  \begin{center}
  \includegraphics[width=8cm]{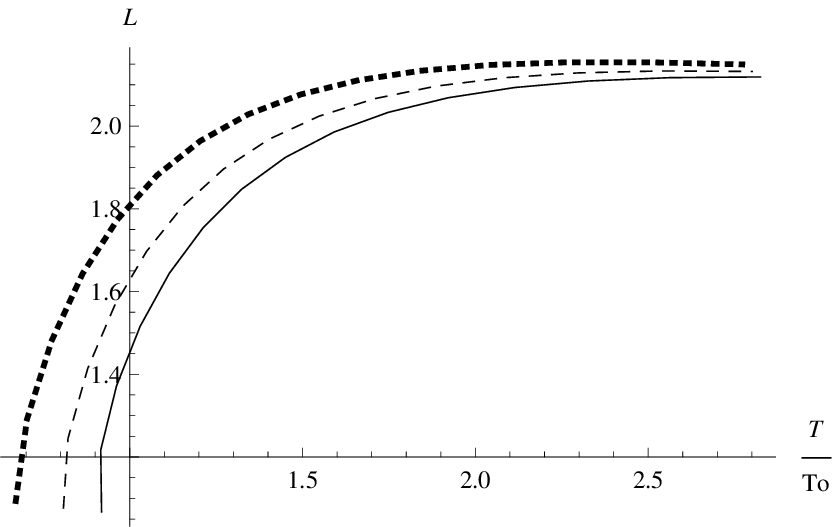}
  \includegraphics[width=8cm]{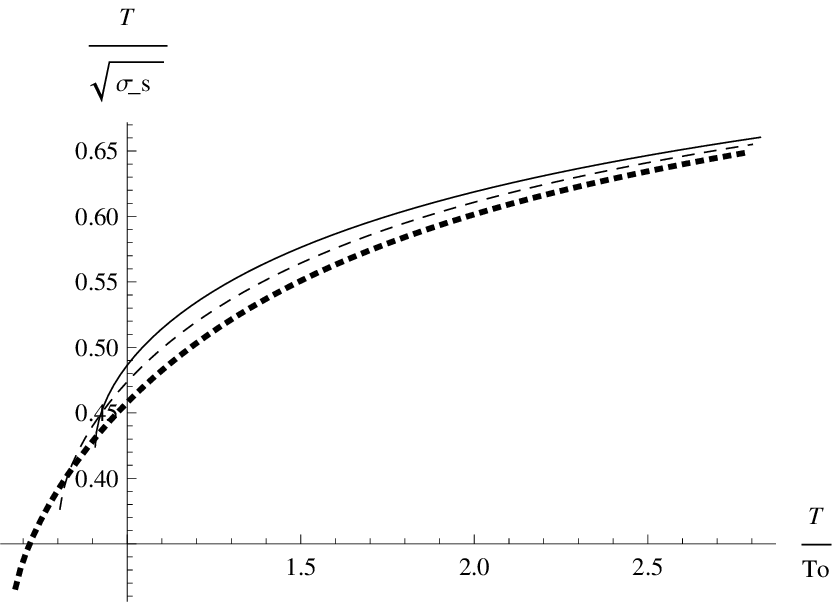}
  \caption{Left: Polyakov line with $\mu / \Lambda=0 \ \text{(solid)}, \ 1 \ \text{(dashed)}, \ 1.5 \ \text{(dotted)}$; ($w_0=1$). Right: Big black hole contribution to the spatial string tension at densities $\mu / \Lambda=0 \ \text{(solid)}, 1.5 \ \text{(dashed)}, 2.5 \ \text{(dotted)}$.}
  \label{figurepolyakovline}
  \end{center}
\end{figure}

The expectation value of the Polyakov line can be schematically written as \cite{Maldacena:1998im, Andreev:2009zk}
\be
\langle L(T,\mu) \rangle= \sum_n w_n Exp(-S_n) \ , \label{polyakovapproximation}
\ee
with weights $w_n$ associated with a renormalized area $S_n$. We will approximate the area using the Nambu-Goto action describing a fundamental string
stretched between the horizon and the boundary at $z=0$. While on-shell quantities such as the susceptibilities do not depend on the frame, quantities such
as the Polyakov line may depend on it. The appropriate background for the string is given by the metric in the string frame
\be
g^s_{AB}(z)=e^{\frac{4}{3} \phi(z)} g_{AB}(z) \ ,
\ee
where $g_{AB}$ is given in eqn. (\ref{metric}).
For a static configuration with the parametrization $\xi_1=t$, $\xi_2=z$ the action is given by
\begin{eqnarray}
S_{NG} &=& \frac{1}{2 \pi \alpha'} \int d^2 \xi \sqrt{g^s_{AB}\partial_M X^A \partial_N X^N} \\
&=& \frac{1}{2 \pi \alpha' T} \int_0^{z_H} dz b(z)^2 \lambda(z)^{\frac{4}{3}} \sqrt{1+f(z) \vec{x} \ '(z)^2}.
\label{ngaction}
\end{eqnarray}
The first integral gives the equation of motion for $\vec{x}(z)$
\be
\frac{d}{dz} \left(\frac{b(z)^2 \lambda(z)^{\frac{4}{3}} f(z) \vec{x} \ '(z)}{\sqrt{1+f(z) \vec{x} \ '(z)^2}} \right)=0.
\ee
It is easy to check that the dominant contribution to the integral in (\ref{ngaction}) comes from the solution $\vec{x}=const.$, and we will neglect other
solutions. Since our metric is asymptotically RN-AdS, implying that $b(z\rightarrow0) = \frac{\mathcal{L}}{z}$, the action defined in (\ref{ngaction})
is divergent. We renormalize the action by subtracting the vacuum contribution stretching from the boundary ($z=0$) upto the wall at $z_w$. The Polyakov line in the saddle-point approximation (\ref{polyakovapproximation}) reads
\begin{eqnarray}
\langle L(T,\mu) \rangle &\simeq& w_0 Exp\left(-S^{ren}_{NG} \right) \\
&=& w_0 Exp\left(\frac{1}{2 \pi \alpha' T} \int_0^{z_H} dz b(z)^2 \lambda(z)^{\frac{4}{3}} -\frac{1}{2 \pi \alpha' T} \int_0^{z_w}dz b(z)^2 \lambda(z)^{\frac{4}{3}} \right)\\
&=&w_0 Exp\left(\frac{1}{2 \pi \alpha' T} \int_{z_w}^{z_H} dz b(z)^2 \lambda(z)^{\frac{4}{3}} \right) .
\end{eqnarray}
At vanishing chemical potential, the 'low' temperature limit of the Polyakov line is given by
\be
\langle L(T\rightarrow T_{min},0) \rangle \propto Exp\Big[ Exp\left(-\frac{1}{Ln(T/T_{min})}\right)\left(\frac{1}{Ln(T/T_{min})}\right)^{\frac{4}{3(q-1)}}\Big]
\ee
and, thus, shows little deviation from its high temperature asymptotic value. Fig. \ref{figurepolyakovline} shows the behavior of the approximate order parameter of the phase transition. The constant $b_0^2 / \alpha'$ is fixed by comparing the spatial string tension with lattice data, see next section. At fixed $\mu$ the expectation value of the Polyakov line jumps rapidly from a minimum value at a temperature $T = T_{min}$ and approaches a constant value in the high temperature region. A direct comparison with lattice data is unfortunately not possible given the subtraction dependence inherent in the definition of the Polyakov line both on the lattice and in our case. This point is generally overlooked in most analyses.

\section{String tensions}
The analyses of the spatial and temporal (effective)  string tension, $\sigma_s$ and $\sigma$ at finite temperature and density, are identical to
those carried in~\cite{Alanen:2009ej} at finite temperature. Indeed, a rerun of their analysis shows that for the spatial string tension $\sigma_s$,
the density dependence is entirely encoded in the position of the horizon $z_H(T, \mu)$ with

\begin{eqnarray}
\sigma_s(T,\mu)=\frac{1}{2 \pi \alpha'}b_s(z)^2 \Big|_{z=z_H}=\frac{1}{2 \pi \alpha'}b(z)^2\lambda(z)^{\frac{4}{3}} \Big|_{z=z_H},
\end{eqnarray}
where $b_s(z)$ is the warping in the string frame.
A comparison with lattice data \cite{Cheng:2008bs} at $T=T_0$, $\mu=0$ fixes $\frac{b_0}{\sqrt{\alpha'}}=3*10^{-4}\text{MeV}^{-1}$. For chemical potentials up to $\mu=2.5 \Lambda$, the spatial string tension  shows a weak dependence on the chemical potential in this improved holographic
model.

\begin{figure}[!htbp]
  \begin{center}
  \includegraphics[width=8cm]{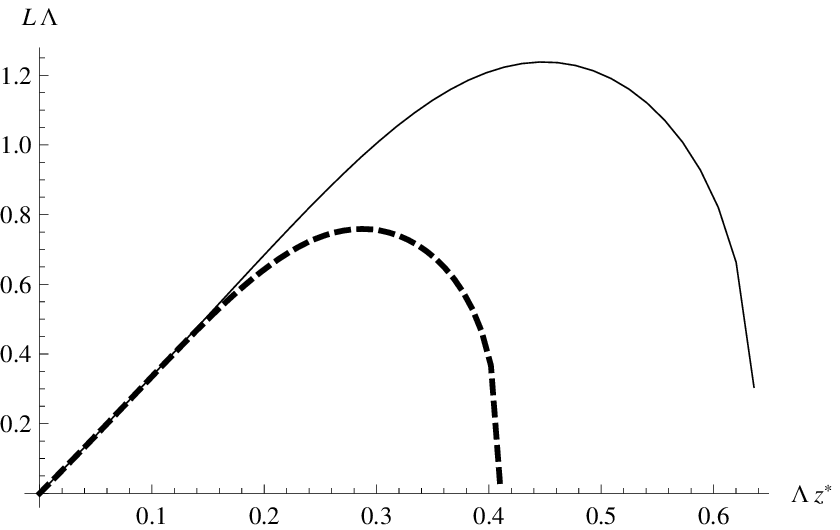}
  \includegraphics[width=8cm]{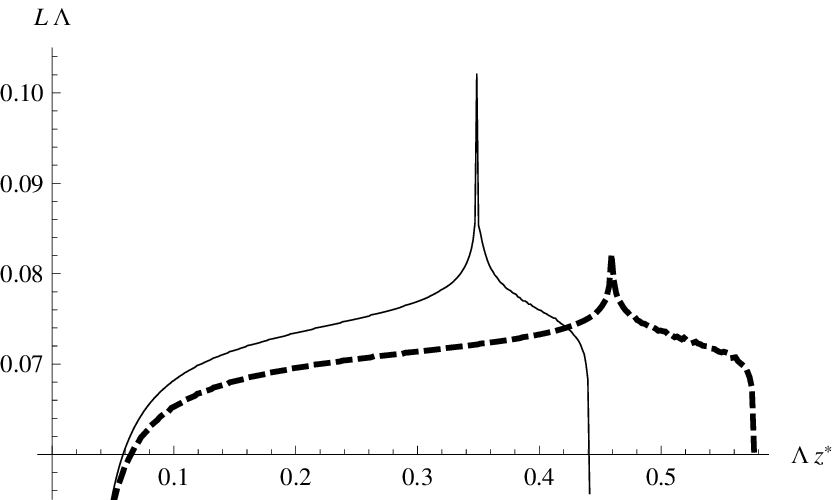}
  \caption{Length separating two heavy quarks. Left: $\mu=0$ with $T=T_0$ (solid) and $T=1.5T_0$ (dashed).  
  Right: $T=T_0, \mu=\mu_0=2.8\Lambda$ (solid) and $T=0, \mu=4 \Lambda$ (dashed).}
  \label{figurelength}
  \end{center}
\end{figure}

In the temporal Wilson loop, the string tying a heavy quark and antiquark in the bifundamental representations
extends in the holographic direction in bulk. The further the spatial separation $L$, the deeper the string 
extends in bulk $z$. Specifically~\cite{Alanen:2009ej}

\begin{eqnarray}
L(z^*)= 2 \int_0^{z^*} dz \frac{1}{\sqrt{f(z)\left(\frac{b_s^4(z) f(z)}{b_s(z^*)f(z^*)}-1 \right)}}
\end{eqnarray}
with $z^*$ the maximum holographic depth for the pending string. In the black hole background
with $f\neq1$, $L(z^*)$ is finite for any temperature and chemical potential as shown  in Fig. \ref{figurelength}.

In the black hole background two heavy-quarks still tie up to distances of order $L_{max}$ which is the 
location of the maximum in Fig.~\ref{figurelength}. In other words,  for $0<L<L_{max}$ we expect a linear-like
free energy between two heavy quarks, while for $L>L_{max}$ the heavy quarks "screen" through the
large entropy of the string (at zero density) or through fundamental matter (at finite density) causing the
free energy to plateau at twice the screening masses. Around the phase transition points, 
$T=T_0$ and $\mu=0$, $L_{max}=1.24\Lambda^{-1}\approx 0.76\,fm$, while for $T=0$ and $\mu=\mu_0=2.8\Lambda$,
$L_{max}=0.10\Lambda^{-1}\approx 0.06\, fm$.  The larger the density, the larger the screening.

\section{Conclusions}

Finite density QCD both at zero and finite temperature is still a challenging problem from first principles.
In the double limit of a large number of colors and large t'Hooft coupling the holographic approach offers
a non-perturbative  tool for investigating QCD-like gauge theories at finite temperature and density. The current
analysis provides a step in that direction whereby finite density effects are incorporated in an improved
AdS/QCD model with a bulk U(1) charge.
Without a dilaton-gauge coupling, our improved AdS/QCD metric at finite (charge) density
is asymptotically RN-AdS on the boundary with a non-trivial dilaton profile interpreted as a running coupling.
The effects of the density do not alter the warping factor of the underlying gravitational metric.

At zero density but for temperatures larger than a minimum temperature, the gravitational equations yield a pair
of black holes of different holographic sizes. A small black hole that is unstable thermodynamically, and a large
black hole that is stable. The occurence of a minimum temperature in the improved model reflects on the (first order)
transition from a 'vacuum' (thermal gas) to a black hole solution. The unstable solution is found to disappear
at large densities. For zero temperature, the small black hole solution disappears while the
large black hole solution requires a minimum chemical potential $\mu_0=2.8\Lambda$. As a result, the metric becomes
AdS$_2 \times \mathbb{R}^3$ in the infrared.  Evaluating the pressure at finite $\mu$ and $T=0$ shows that
the 'vacuum' becomes unstable against black hole solution  at a critical $\mu=\mu_0$ indicating a first order
Hawking-Page transition.

The (quark) susceptibilities $c_2, \ c_4, \ c_6$ show rapid variations across the critical temperature $T_0$
in the improved holographic model, in contrast to the RN-AdS model where they are found to be constant.
The variations are overall consistent with the ones reported by lattice simulations. However, their
asymptotic ratios are off compared to the free fermion (quark) limits, despite the fact that in the improved
holographic model the gauge coupling on the boundary runs weak at high temperature.

To further clarify the nature of the charged black hole solutions, we have analyzed the subtracted Polyakov 
line and found that it jumps at both finite temperature and/or finite density. The jump is suggestive of 
screening which is expected to be larger at finite charge density.  Indeed, our analysis of the temporal
Wilson loop confims that two heavy quarks detach more easily at higher density. Spatial Wilson loops are
midly affected by the charge screening in the black-hole background.

The bottom-up model with running coupling constant appears to capture some essentials of a first order
transition from a vacuum with matter characterized by $f=1$ and a charged black-hole characterized by
$f\neq 1$. This first order  geometrical transition captured by the holographic gravity equations appears to 
encode some of the features expected from a first order transition in QCD with a large number of colors
in the homogeneous regime. The latter is expected to be dominant for a broad range of temperatures and
densities. However, it has two major shortcomings: 1/ The (quark) susceptibilities asymptote the wrong
values despite the fact that the bulk thermodynamics (pressure, entropy and energy densities) can be
adjusted to asymptote the correct values. 2/ The very low temperature phase is likely inhomogeneous.
While the latter issue is readily overcome~\cite{HOLO}, the former issue is more problematic 
at higher temperature. It goes to the heart of high fermionic charge fluctuations in dense matter
and therfore the very interpretation of our charged black hole solutions.

\vskip 1in
{\bf Acknowledgements}
\vskip 1cm
\noindent This work was supported in part by US-DOE grants
DE-FG02-88ER40388 and DE-FG03-97ER4014.

\newpage

\appendix

\section{Non-trivial dilaton-gauge coupling: $c=4$}

In this Appendix, we show that a non-trivial dilaton-gauge coupling affects considerably the
formation of the large black hole solution in the model we discussed above.  

\begin{figure}[!htbp]
  \begin{center}
  \includegraphics[width=7cm]{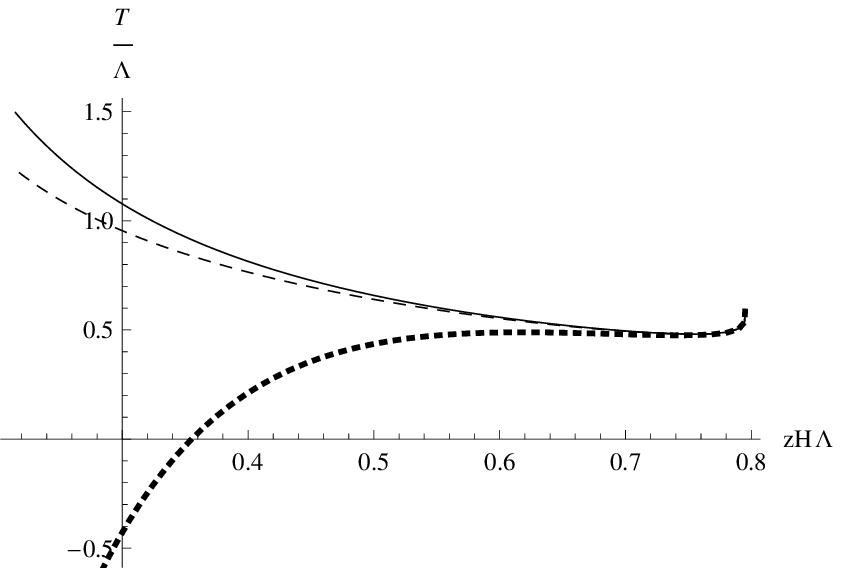}
  \includegraphics[width=7cm]{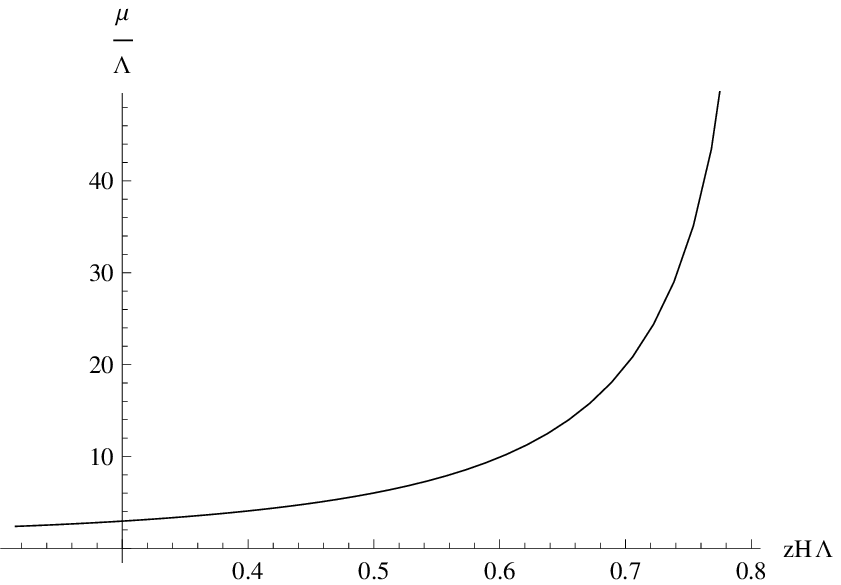}
  \caption{Left: The Hawking temperature at chemical potentials $\mu / \Lambda=0 \ \text{(solid)}, \ 1 \ \text{(dashed)}$, \ 3.5 \ \text{(dotted)} as a function of the scaled horizon $z_H \Lambda$. Right: The density at $T=0$.}
  \label{figuretemperaturec4}
  \end{center}
\end{figure}

Fig.~\ref{figuretemperaturec4} shows, that the big black hole branch vanishes completely for densities larger than $\mu_0$ leaving only the unstable solution with an absolute maximum temperature in the IR region ($z_H \rightarrow z_w$). In this region, the size of the black hole at a given temperature is insensitive to changes in the charge density. Discarding the unstable small black hole, the range of validity for this setup is dictated by $T \geq T_{min}$, $\mu \leq \mu_0$.

\begin{figure}[!htbp]
  \begin{center}
  \includegraphics[width=8cm]{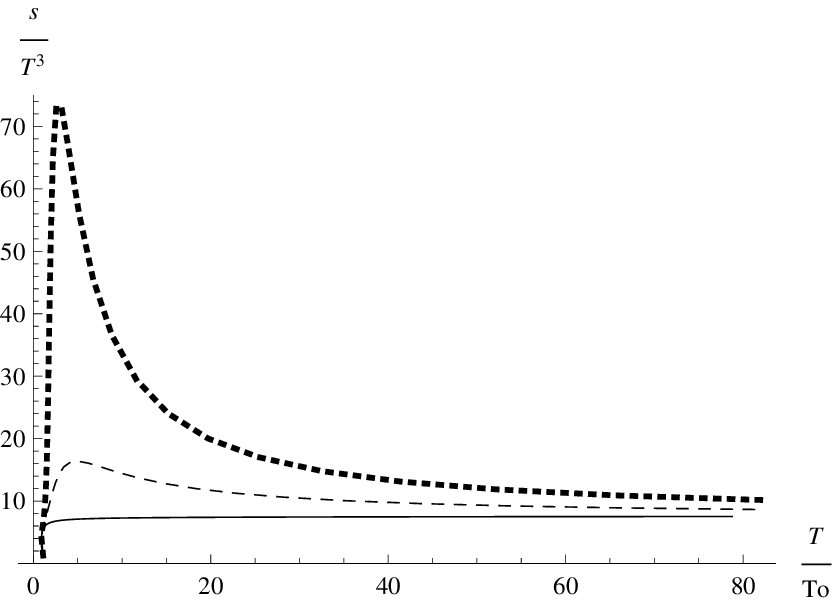}
  \includegraphics[width=8cm]{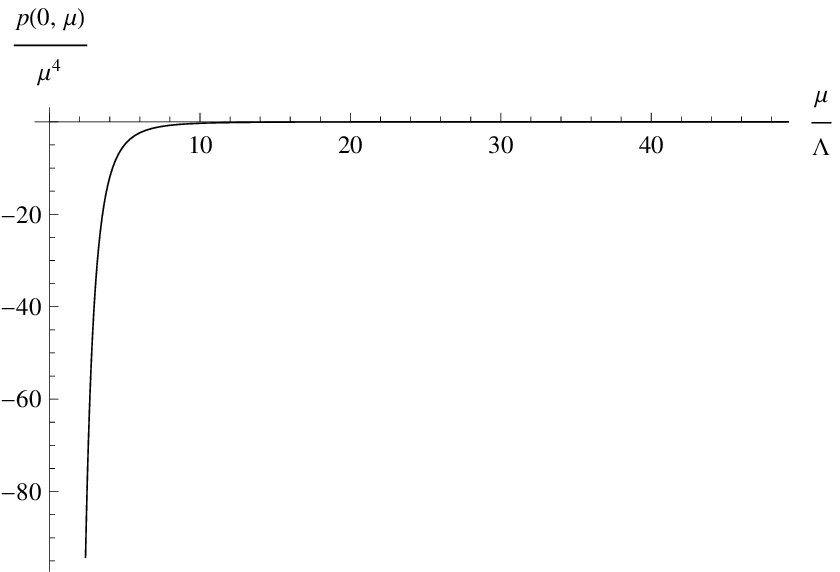}
  \caption{The scaled entropy density $s/T^3$ at chemical potentials $\mu / \Lambda=0 \ \text{(solid)}, \ 1 \ \text{(dashed)}$, \ 1.5 \ \text{(dotted)} and the pressure at $T=0$.}
  \label{figureentropyc4}
  \end{center}
\end{figure}

The scaled entropy density, $s/T^3$, peaks at finite $\mu$ around $T=5 T_0$ and the high temperature asymptotic behavior is much slower as in the case of vanishing dilaton-gauge coupling, Fig. \ref{figureentropyc4}. The susceptibilities $c_4, c_6$ do not asymptote for large temperatures. The pressure at $T=0$ is negative for small $\mu / \Lambda$ and asymptotes to zero for $\mu/ \Lambda \ge 10$. Thus, the 'vacuum' solution is stable against the black hole solution at $T=0$ for all ranges of the chemical potential and no phase transition expected.

\begin{figure}[!htbp]
  \begin{center}
  \includegraphics[width=8cm]{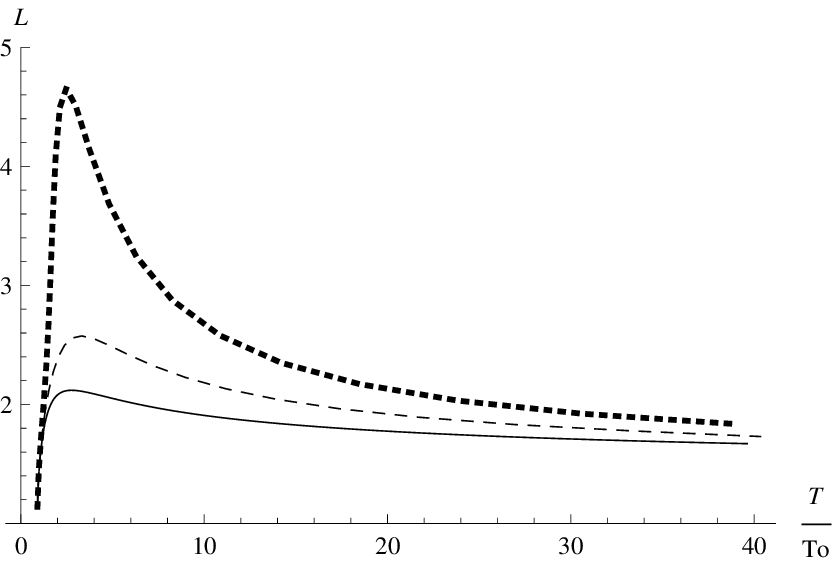}
  \includegraphics[width=8cm]{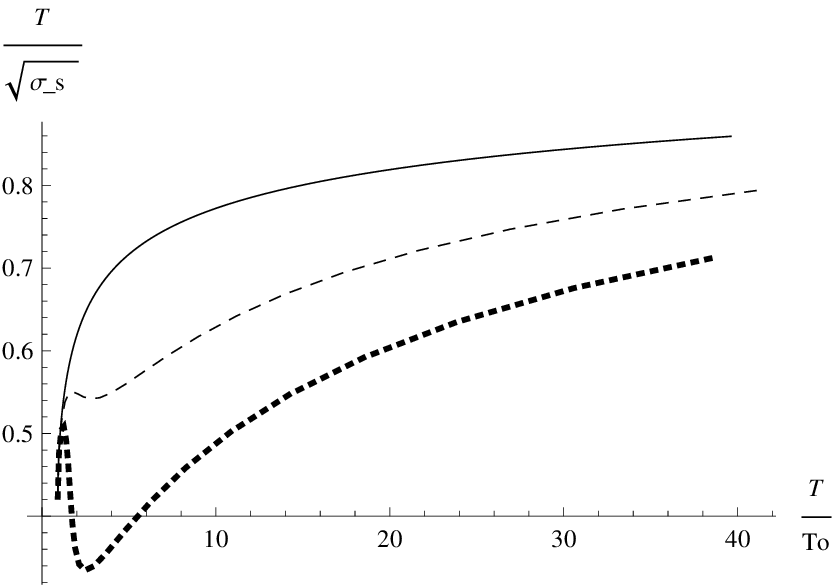}
  \caption{Left: Polyakov line with $\mu / \Lambda=0 \ \text{(solid)}, \ 1 \ \text{(dashed)}, \ 1.5 \ \text{(dotted)}$. Right: Big black hole contribution to the spatial string tension at densities $\mu / \Lambda=0 \ \text{(solid)}, 1.5 \ \text{(dashed)}, 2.5 \ \text{(dotted)}$. ($w_0=1$)}
  \label{figuresigmasc4}
  \end{center}
\end{figure}

The Polyakov line shows a peak around $T=T_0$ that is enhanced with increasing charge density. Fig. \ref{figuresigmasc4} shows the results for the Polyakov line and the spatial string tension with non-trivial dilaton-gauge coupling.

\section{Warped Model}

We discuss the asymptotics of the susceptibilites as obtained for the warped metric proposed in~\cite{Andreev:2006nw}
with a charged black hole. At finite temperature and density the metric reads
\be
ds^2=\frac{\mathcal{L}^2}{z^2} e^{0.45\text{GeV}^2 z^2} \left(-f(z) dt^2+d\vec{x} \ ^2+\frac{1}{f(z)}dz^2 \right)
\ee
and we assume that the relation between the horizon $z_H$ and the temperature, density is given by the RN-AdS result \cite{Sin:2007ze}
\be
z_H=2 \left(\pi T + \sqrt{\pi^2 T^2 + \frac{4}{3}\gamma^2 \mu^2} \right)^{-1}. \label{temperatureandreev}
\ee
The solution (\ref{electrostaticpotential}) is generic and we obtain
\be
\mu = \frac{\textbf{e}}{\mathcal{L}^3} \int_0^{z_H}dx \frac{x}{\mathcal{L}} e^{-0.225\text{GeV}^2 x^2}=-\frac{\textbf{e}}{\mathcal{L}^4 0.45\text{GeV}^2} e^{0.225\text{GeV}^2z_H^2}. \label{densityandreev}
\ee
The charge density is proportional to the charge and using (\ref{densityandreev}) yields
\be
n(T,\mu) \propto \mu e^{0.225\text{GeV}^2z_H^2}.
\ee
Unlike the improved holographic model, the corresponding susceptibilities vanish as power laws at high temperature,
\begin{eqnarray}
c_2 \propto \frac{1}{T^2} e^{ \frac{0.225\text{GeV}^2}{\pi^2 T^2}} \rightarrow 0 \\
c_4 \propto \frac{1}{T^4} e^{ \frac{0.225\text{GeV}^2}{\pi^2 T^2}} \rightarrow 0\\
c_6 \propto \left( \# \frac{1}{T^6} + \# \frac{1}{T^4} \right) e^{ \frac{0.225\text{GeV}^2}{\pi^2 T^2}}\rightarrow 0
\end{eqnarray}
This warped and charged black hole cannot be used to model up the fermionic fluctuations in screened QCD.

\end{document}